\documentclass[twocolumn,showpacs,aps,prl,superscriptaddress]{revtex4}

\usepackage{graphicx}
\usepackage{dcolumn}
\usepackage{amsmath}
\usepackage{epsfig}

\newcommand{\BABARPubYear}    {03}
\newcommand{\BABARPubNumber}  {013}

\newcommand{\SLACPubNumber}   {9923}
\newcommand{\LANLNumber}      {0306030}

\RequirePackage{xspace}

\usepackage{relsize}
\def\babar{\mbox{\slshape B\kern-0.1em{\smaller A}\kern-0.1em
    B\kern-0.1em{\smaller A\kern-0.2em R}}}

\def\epem       {\ensuremath{e^+e^-}\xspace}
\def\ee         {\ensuremath{e^-e^-}\xspace}

\def\pip   {\ensuremath{\pi^+}\xspace}
\def\pim   {\ensuremath{\pi^-}\xspace}

\def\Kbar  {\kern 0.2em\overline{\kern -0.2em K}{}\xspace}

\def\Kz    {\ensuremath{K^0}\xspace}
\def\Kzb   {\ensuremath{\Kbar^0}\xspace}
\def\KzKzb {\ensuremath{\Kz \kern -0.16em \Kzb}\xspace}
\def\Kp    {\ensuremath{K^+}\xspace}
\def\Km    {\ensuremath{K^-}\xspace}

\def\KpKm  {\ensuremath{\Kp \kern -0.16em \Km}\xspace}

\def\Dbar    {\kern 0.2em\overline{\kern -0.2em D}{}\xspace}

\def\Dz      {\ensuremath{D^0}\xspace}
\def\Dzb     {\ensuremath{\Dbar^0}\xspace}
\def\DzDzb   {\ensuremath{\Dz {\kern -0.16em \Dzb}}\xspace}
\def\Dp      {\ensuremath{D^+}\xspace}
\def\Dm      {\ensuremath{D^-}\xspace}

\def\DpDm    {\ensuremath{\Dp {\kern -0.16em \Dm}}\xspace}

\def\B       {\ensuremath{B}\xspace}
\def\Bbar    {\kern 0.18em\overline{\kern -0.18em B}{}\xspace}

\def\Bz      {\ensuremath{B^0}\xspace}
\def\Bzb     {\ensuremath{\Bbar^0}\xspace}
\def\BzBzb   {\ensuremath{\Bz {\kern -0.16em \Bzb}}\xspace}
\def\Bu      {\ensuremath{B^+}\xspace}
\def\Bub     {\ensuremath{B^-}\xspace}

\def\BpBm    {\ensuremath{\Bu {\kern -0.16em \Bub}}\xspace}

\def\BorBbar    {\kern 0.18em\optbar{\kern -0.18em B}{}\xspace}
\def\DorDbar    {\kern 0.18em\optbar{\kern -0.18em D}{}\xspace}
\def\KorKbar    {\kern 0.18em\optbar{\kern -0.18em K}{}\xspace}

\mathchardef\Upsilon="7107
\def\Y#1S{\ensuremath{\Upsilon{(#1S)}}\xspace}

\def\FourS {\Y4S}

\mathchardef\Deltares="7101
\mathchardef\Xi="7104
\mathchardef\Lambda="7103
\mathchardef\Sigma="7106
\mathchardef\Omega="710A

\def\Deltabar{\kern 0.25em\overline{\kern -0.25em \Deltares}{}\xspace}
\def\Lbar{\kern 0.2em\overline{\kern -0.2em\Lambda\kern 0.05em}\kern-0.05em{}\xspace}
\def\Sigbar{\kern 0.2em\overline{\kern -0.2em \Sigma}{}\xspace}
\def\Xibar{\kern 0.2em\overline{\kern -0.2em \Xi}{}\xspace}
\def\Obar{\kern 0.2em\overline{\kern -0.2em \Omega}{}\xspace}
\def\Nbar{\kern 0.2em\overline{\kern -0.2em N}{}\xspace}
\def\Xb{\kern 0.2em\overline{\kern -0.2em X}{}\xspace}

\def\mes        {\mbox{$m_{\rm ES}$}\xspace}

\newcommand{\tev}{\ensuremath{\mathrm{\,Te\kern -0.1em V}}\xspace}
\newcommand{\gev}{\ensuremath{\mathrm{\,Ge\kern -0.1em V}}\xspace}
\newcommand{\mev}{\ensuremath{\mathrm{\,Me\kern -0.1em V}}\xspace}
\newcommand{\kev}{\ensuremath{\mathrm{\,ke\kern -0.1em V}}\xspace}
\newcommand{\ev}{\ensuremath{\mathrm{\,e\kern -0.1em V}}\xspace}
\newcommand{\gevc}{\ensuremath{{\mathrm{\,Ge\kern -0.1em V\!/}c}}\xspace}
\newcommand{\mevc}{\ensuremath{{\mathrm{\,Me\kern -0.1em V\!/}c}}\xspace}
\newcommand{\gevcc}{\ensuremath{{\mathrm{\,Ge\kern -0.1em V\!/}c^2}}\xspace}
\newcommand{\mevcc}{\ensuremath{{\mathrm{\,Me\kern -0.1em V\!/}c^2}}\xspace}

\def\invfb   {\ensuremath{\mbox{\,fb}^{-1}}\xspace}

\def\mus  {\ensuremath{\rm \,\mus}\xspace}

\def\ps   {\ensuremath{\rm \,ps}\xspace}

\def\mus        {\ensuremath{\,\mu{\rm s}}\xspace}    
\def\ps         {\ensuremath{{\rm \,ps}}\xspace}  

\def\to                 {\ensuremath{\rightarrow}\xspace}

\def\pep2{PEP-II}

\def\gsim{{~\raise.15em\hbox{$>$}\kern-.85em
          \lower.35em\hbox{$\sim$}~}\xspace}
\def\lsim{{~\raise.15em\hbox{$<$}\kern-.85em
          \lower.35em\hbox{$\sim$}~}\xspace}

\def\CP                {\ensuremath{C\!P}\xspace}

\def\deltat{\ensuremath{{\rm \Delta}t}\xspace}
\def\deltamd{\ensuremath{{\rm \Delta}m_d}\xspace}

\xspace

\newcommand{\jprlBase}       {Phys.\ Rev.\ Lett.\xspace}

\newcommand{\jprl}      [1]  {\jprlBase\ {\bf #1}}

\newcommand{\progtp}    [1]  {{Prog.\ Th.\ Phys.\ {\bf #1}}}

\def\jetset74   {\mbox{\tt Jetset \hspace{-0.5em}7.\hspace{-0.2em}4}\xspace}

\newcommand{\bit}{\begin{itemize}}
\newcommand{\eit}{\end{itemize}}
\newcommand{\beq}{\begin{equation}}
\newcommand{\eeq}{\end{equation}}
\newcommand{\beqn}{\begin{eqnarray}}
\newcommand{\eeqn}{\end{eqnarray}}
\newcommand{\beqns}{\begin{eqnarray*}}
\newcommand{\eeqns}{\end{eqnarray*}}

\newcommand{\mc}{\multicolumn}

\def\rs{\raisebox{1.5ex}[-1.5ex]}

\def\rar{\rightarrow}

\def\rhopi{\rho\pi}
\def\rhoX{\rho h}
\def\rhok{\rho K}

\def\AcprhoX{A_{CP}^{\rhoX}}
\def\Acprhopi{A_{CP}^{\rhopi}}
\def\AcprhoK{A_{CP}^{\rho K}}

\def\hel{{\theta_{\pi}}}
\def\cshelrho{{\rm cos}\,\hel}

\def\de{\Delta E}
\def\mes{\ensuremath{m_{\rm ES}}}

\def\cat{k}
\def\cont{{q\bar q}}
\def\MeVc2{${\rm MeV}/c^2$}
\def\GeVc2{${\rm GeV}/c^2$}

\def\babar{$\mbox{\sl B\hspace{-0.4em} {\small\sl A}\hspace{-0.37em} 
\sl B\hspace{-0.4em} {\small\sl A\hspace{-0.02em}R}}$}

\def\ee{$e^+e^-$}

\def\NN{{\rm NN}}

\def\Amp{A_{\rho\pi}}
\def\Ampbar{\kern 0.18em\overline{\kern -0.18em A}{}_{\rho\pi}}

\def\ee{\ensuremath{e^+e^-}}
\def\corr{c}
\def\eg{{\em e.g.}}

\def\Nbar{\overline{N}}

\def\mes {\ensuremath{m_{ES}}\xspace}

\def\Dbar{\ensuremath{\overline{D}}\xspace}

\def\dm {\ensuremath{\Delta m_d}\xspace}
\def\dt {\ensuremath{\Delta t}}

\def\dC{\ensuremath{\Delta C}\xspace}
\def\dS{\ensuremath{\Delta S}\xspace}

\def\figurebox#1#2#3{%
    \def\arg{#3}%
    \ifx\arg\empty
    {\hfill\vbox{\hsize#2\hrule\hbox to #2{\vrule\hfill\vbox to #1{\hsize#2\vfill}\vrule}\hrule}\hfill}%
    \else
    {\hfill\epsfbox{#3}\hfill}%
    \fi}

\begin{document}

\preprint{\babar-PUB-\BABARPubYear/\BABARPubNumber} 
\preprint{SLAC-PUB-\SLACPubNumber} 

\begin{flushleft}
\babar-PUB-\BABARPubYear/\BABARPubNumber\\
SLAC-PUB-\SLACPubNumber\\
hep-ex/\LANLNumber\\[10mm]
\end{flushleft}

\title{
{\large \bf \boldmath
Measurements of Branching Fractions and 
{\em CP}-Violating Asymmetries \\
in \boldmath$B^0\to\rho^\pm h^\mp$ Decays} 
}

%
\author{B.~Aubert}
\author{R.~Barate}
\author{D.~Boutigny}
\author{J.-M.~Gaillard}
\author{A.~Hicheur}
\author{Y.~Karyotakis}
\author{J.~P.~Lees}
\author{P.~Robbe}
\author{V.~Tisserand}
\author{A.~Zghiche}
\affiliation{Laboratoire de Physique des Particules, F-74941 Annecy-le-Vieux, France }
% ...
\collaboration{The \babar\ Collaboration}
\noaffiliation

\date{\today}

\begin{abstract}
We present measurements of branching fractions and 
\CP-violating asymmetries in $B^0\rar\rho^{\pm}\pi^{\mp}$ and 
$B^0\rar\rho^{-} K^{+}$ decays. The results are obtained from a 
data sample of $88.9 \times 10^6$ $\FourS \to B\Bbar$ decays
collected with the \babar\ detector at the \pep2 asymmetric-energy 
$B$~Factory at SLAC. From a time-dependent maximum likelihood fit 
we measure the charge-averaged branching fractions
${\cal B}(\Bz\to\rho^{\pm}\pi^{\mp}) = (22.6 \pm 1.8 {\rm\,(stat)}
	\pm2.2{\rm\,(syst)})\times 10^{-6}$ and 
${\cal B}(\Bz\to\rho^{-}K^{+}) = (7.3^{\,+1.3}_{\,-1.2}
	\pm1.3)\times 10^{-6}$;
and the \CP-violating charge asymmetries  
$\Acprhopi=-0.18\pm 0.08\pm{0.03}$
and 
$\AcprhoK=0.28\pm 0.17 \pm{0.08}$;
the direct \CP\ violation parameter
$C_{\rho\pi}=0.36\pm 0.18\pm{0.04}$
and the mixing-induced \CP\ violation parameter
$S_{\rho\pi}=0.19\pm 0.24 \pm{0.03}$; and
the dilution parameters 
$\dC_{\rho\pi}=0.28^{\,+0.18}_{\,-0.19} \pm{0.04}$
and $\dS_{\rho\pi}=0.15\pm 0.25\pm{0.03}$. 
\end{abstract}


\maketitle

In the Standard Model, $\CP$-violating effects arise from a single
complex phase in the three-generation 
Cabibbo-Kobayashi-Maskawa quark-mixing matrix~\cite{CKM}.  
One of the central, unresolved  questions 
is whether this
mechanism is sufficient to explain the pattern of \CP violation observed 
in nature. 
We present here a simultaneous measurement of branching fractions 
and \CP-violating asymmetries in the decays $B^0\rar\rho^{\pm}\pi^{\mp}$ 
and $B^0\rar\rho^{-} K^{+}$ (and their charge conjugates).
The \babar\ 
and Belle experiments have performed searches 
for \CP-violating asymmetries 
in $B$ decays to $\pip\pim$~\cite{bib:BabarSin2alpha,bib:BelleSin2alpha},
where the mixing-induced \CP
asymmetry is related to the angle 
$\alpha \equiv \arg\left[-V_{td}^{}V_{tb}^{*}/V_{ud}^{}V_{ub}^{*}\right]$ 
of the Unitarity Triangle as it is for $\rho^\pm\pi^\mp$.
However, unlike $\pip\pim$,
$\rho^{\pm}\pi^{\mp}$ is not a \CP eigenstate, and four flavor-charge
configurations
$(\Bz(\Bzb) \to \rho^{\pm}\pi^{\mp})$ must be considered.  
Although
this leads to a more complicated analysis, it
benefits from a branching fraction that is nearly five times 
larger~\cite{bib:BaBarRhopi,bib:BRRhopi}. 

Following a quasi-two-body approach~\cite{bib:BaBarPhysBook},
we restrict the analysis to the two regions of the 
$\pi^{\mp}\pi^0 h^{\pm}$ Dalitz plot ($h=\pi$ or $K$) that are dominated by
either~$\rho^{+}h^{-}$ or~$\rho^{-}h^{+}$. 
With $\deltat \equiv t_{\rhoX} - t_{\rm tag}$ defined as the proper time interval 
between the decay of the reconstructed $B^0_{\rhoX}$ and that of the 
other meson $\Bz_{\rm tag}$,  the time-dependent decay rates are given by 
\beqn
\label{eq:thTime}
  \lefteqn{f^{\rho^\pm h^\mp}_{Q_{\rm tag}}(\deltat) = (1\pm \AcprhoX)
           \frac{e^{-\left|\deltat\right|/\tau}}{4\tau}} \\
	&&\hspace{1.3cm}\times\,\bigg[1+Q_{\rm tag} 
             (S_{\rho h}\pm\dS_{\rho h})\sin(\deltamd\deltat)\nonumber\\[-0.1cm]
	&&\hspace{1.3cm}\phantom{\times\,\bigg[1}
	    -Q_{\rm tag} 
		(C_{\rho h}\pm\dC_{\rho h})\cos(\deltamd\deltat)\bigg]\;,\nonumber
\eeqn
where $Q_{\rm tag}= 1(-1)$ when the tagging meson $\Bz_{\rm tag}$
is a $\Bz(\Bzb)$, $\tau$ is the mean 
\Bz lifetime, and $\deltamd$ is the $\BzBzb$ oscillation frequency.
The time- and flavor-integrated 
charge asymmetries $\Acprhopi$ and $\AcprhoK$
measure direct \CP violation.  
For the $\rhopi$ mode, the quantities $S_{\rho\pi}$ and $C_{\rho\pi}$ 
parameterize mixing-induced \CP violation related to the angle $\alpha$,
and flavor-dependent direct \CP violation, respectively.
The parameters $\dC_{\rho \pi}$ and $\dS_{\rho\pi}$ are insensitive to
\CP violation. 
$\dC_{\rho \pi}$ describes 
the asymmetry between the rates 
$\Gamma({\Bz} \to{\rho^+\pi^-}) + \Gamma({\Bzb} \to {\rho^-\pi^+})$ and
${\Gamma(\Bz} \to {\rho^-\pi^+}) + \Gamma({\Bzb} \to {\rho^+\pi^-})$, while
$\dS_{\rho\pi}$ is related to the strong phase difference between
the amplitudes contributing to $\Bz \to\rhopi$ decays. More precisely,
one finds the relations 
$S_{\rho \pi}\pm\Delta S_{\rho \pi} = 
\sqrt{1-(C_{\rho \pi}\pm\dC_{\rho \pi})^2}\,\sin(2\alpha_{\rm eff}^{\pm}\pm\delta)$,
where $2\alpha_{\rm eff}^\pm=\arg[(q/p)(\Ampbar^{\pm}/\Amp^{\mp})]$,
$\delta=\arg[\Amp^{-}/\Amp^{+}]$, 
$\arg[q/p]$ is the $\BzBzb$ mixing phase,
and $\Amp^+(\Ampbar^+)$ and 
$\Amp^-(\Ampbar^-)$ are the transition 
amplitudes of the processes $B^0(\Bzb)\to\rho^+\pi^-$ and $B^0(\Bzb)\to\rho^-\pi^+$,
respectively. The angles $\alpha_{\rm eff}^\pm$
are equal to $\alpha$ in the absence of contributions 
from penguin amplitudes. For the self-tagging $\rhok$
mode, the values of the four time-dependent parameters are $C_{\rho K}
= 0$, $\dC_{\rho K} = -1$, $S_{\rho K} = 0$, and $\dS_{\rho K} = 0$.

The data used in this analysis were accumulated 
with the \babar\ detector~\cite{bib:babarNim}, 
at the \pep2 asymmetric-energy $e^+e^-$ storage ring at SLAC.
The sample consists of $(88.9\pm1.0)\times10^{6}$ $B\Bbar$ pairs
collected at the \FourS resonance (``on-resonance''),
and an integrated luminosity of 9.6~\invfb collected about 
40~\mev below the~\FourS (``off-resonance'').
In Ref.~\cite{bib:babarNim} we describe the silicon
vertex tracker and drift chamber used for track and vertex
reconstruction, the Cherenkov detector (DIRC), 
the electromagnetic calorimeter (EMC),
and their use in particle identification (PID).

We reconstruct $B^0_{\rhoX}$ candidates from combinations 
of two  tracks and a $\pi^0$~candidate.
We require that the PID of both tracks be inconsistent with the 
electron hypothesis, and the PID of the track used to form the
$\rho$ be inconsistent with the kaon hypothesis.
The $\pi^0$ candidate mass must satisfy 
$0.11<m(\gamma\gamma)<0.16~\gevcc$, where each photon 
is required to have an energy greater than $50\mev$
in the laboratory frame and to
exhibit a lateral profile of energy deposition in the EMC consistent 
with an electromagnetic shower.
The mass of the $\rho$ candidate must satisfy 
$0.4<m(\pi^\pm\pi^0)<1.3~\gevcc$. To avoid the interference region, 
the $B$~candidate is rejected if both 
the $\pi^+\pi^0$ and $\pi^-\pi^0$ pairs satisfy this requirement.
Taking advantage of the helicity structure of $B\to\rhoX$ decays
($h$ is denoted {\em bachelor track} hereafter), we 
require $|\cshelrho| > 0.25$, 
where $\hel$ is the angle between the $\pi^0$ momentum
and the negative $B$ momentum in the $\rho$ rest frame.
The bachelor track from the $\rho h$ decay 
must have a $\ee$ center-of-mass (CM) 
momentum above~$2.4 \gevc$. 

For $86\%$ of the  $\Bz \to \rho h$ decays  that pass the event selection, 
the pion from the $\rho$ has momentum below this value, and thus the	
charge of the $\rho$ is determined unambiguously. 
For the remaining events, the charge of the $\rho$ is taken to be that 
of the $\pi^{\pm}\pi^0$ combination with mass closer to the 
$\rho$ mass.
With this procedure, $5\%$ of 
the selected simulated signal events are assigned an incorrect 
charge.

To reject background from two-body $B$ decays, the invariant masses 
of the $\pi^\pm h^\mp$ and $h^\pm\pi^0$ combinations must each 
be less than $5.14~\gevcc$.
Two kinematic variables allow the discrimination of signal-$B$ 
decays from fake-$B$ decays 
due to random combinations of tracks and $\pi^0$~candidates. 
One variable is the difference, $\de$, between the CM
energy of the $B$~candidate and $\sqrt{s}/2$, where $\sqrt{s}$ is the 
total CM energy. The other variable is the beam-energy-substituted mass 
$\mes\equiv\sqrt{(s/2+{\mathbf {p}}_i\cdot{\mathbf{p}}_B)^2/E_i^2-{\mathbf {p}}_B^2},$
where the $B$ momentum ${\mathbf {p}}_B$ and the four-momentum of the 
initial state 
($E_i$, ${\mathbf {p}}_i$) are defined in the laboratory frame.
The $\de$ distribution for $\rhopi$ ($\rho K$) signal peaks around 
0 ($-45$)~MeV 
since the pion mass is always assigned to the bachelor track.
We require $5.23 < \mes <5.29~\gevcc$ and $-0.12<\de<0.15~\gev$, where the 
asymmetric $\de$ window suppresses higher-multiplicity $B$ background,
which leads to mostly negative $\de$ values. Discrimination between 
$\rho\pi$ and $\rhok$ events is provided by the Cherenkov
angle $\theta_C$ and, to a lesser extent, by $\de$.

Continuum $e^+e^-\to q\bar{q}$ ($q = u,d,s,c$) events are the
dominant background.
To enhance
discrimination between signal and continuum, we use a
neural network (NN) to combine four discriminating variables: the 
reconstructed $\rho$ mass,  $\cshelrho$, 
and the two event-shape variables that are used in the Fisher
discriminant of Ref.~\cite{bib:BabarSin2alpha}.
The NN is trained in the signal region with off-resonance data and
simulated signal events. The final sample of signal candidates 
is selected with a cut on the NN output that retains $\sim 65\%$ ($5\%$)
of the signal (continuum).

Approximately $23\%$ $(20\%)$ of simulated $\rho\pi$ ($\rho K$) 
events have more than one $\rho h$ candidate passing the 
selection criteria.  
In these cases, we choose the candidate with the reconstructed $\pi^0$ 
mass closest to the nominal $\pi^0$ mass.
A total of 20,497 events pass all selection criteria.
The signal efficiency determined from Monte Carlo 
(MC) simulation is $20.7\%$ ($18.5\%$) for $\rho\pi$ ($\rho K$) events; 
$31\%$ ($30\%$) of the selected events are misreconstructed, mostly 
due to combinatorial-$\pi^0$ background.

We use  MC-simulated events to study the cross-feed 
from other $B$~decays.
The charmless modes are grouped into eleven classes with
similar kinematic and topological properties.
Two additional classes account for the neutral 
and charged $b \to c$ decays. For each of the background classes, 
a component is introduced into the likelihood, with a fixed 
number of events. In the selected $\rhopi$ ($\rhok$) samples
we expect 
$6\pm1$ ($20\pm2$) charmless two-body background events,
$93\pm23$ ($87\pm22$) charmless three-body background events, 
$118\pm65$ ($36\pm18$) charmless four-body background events, and
$266\pm43$ ($54\pm11$) $b \to c$ events.
Backgrounds from two-, three-, and four-body decays to $\rhopi$ are 
dominated by $B^+\to\pi^+\pi^0$, $B^+\to\rho^0\pi^+$, 
and longitudinally polarized $B^0\to\rho^+\rho^-$ decays.
The $\rhok$ sample receives its dominant two-body
background from $B^+\to K^+\pi^0$, and its dominant 
three- and four-body background
from $B\to K{^\ast}\pi$ and higher kaon resonances, estimated from 
inclusive $B\to K\pi\pi$ measurements.


The time difference $\deltat$ is obtained from the measured distance between 
the $z$ positions (along the beam direction) of the $\Bz_{\rhoX}$ and 
$\Bz_{\rm tag}$ decay vertices, and the boost $\beta\gamma=0.56$ of 
the \epem\  system~\cite{bib:BabarS2b,bib:BabarSin2alpha}. 
To determine the flavor of the $\Bz_{\rm tag}$ 
we use the tagging algorithm of Ref.~\cite{bib:BabarS2b}.
This produces four mutually exclusive tagging categories. We also 
retain untagged events in a fifth category to improve the efficiency 
of the signal selection and the sensitivity to charge asymmetries.
Correlations between the $B$ flavor tag and the charge   
of the reconstructed $\rhoX$ candidate  are observed in various 
$B$-background channels and evaluated with MC simulation.

We use an unbinned extended maximum likelihood fit to extract
the $\rhopi$ and $\rhok$ event yields, the \CP parameters and the 
other parameters defined in Eq.~(\ref{eq:thTime}).
The likelihood for the $N_\cat$ candidates $i$ tagged in category 
$k$ is
\begin{equation}
\label{eq:pdfsum}
{\cal L}_k = e^{-N^{\prime}_\cat}\!\prod_{i=1}^{N_\cat}
\sum_{h}^{\pi,K}\bigg\{ N^{\rho h} 
\epsilon_\cat {\cal P}_{i,\cat}^{\rho h}
+ N_\cat^{\cont, h} {\cal P}_{i,\cat}^{\cont, h} 
 + \sum_{j=1}^{N_B} {\cal L}^{\B, h}_{ij, \cat}\bigg\}
\end{equation}
where $N^{\prime}_\cat$ is the sum of the signal and continuum yields
(to be determined by the fit) and the fixed $B$-background yields,
$N^{\rho h}$ is the number of signal events of type ${\rho h}$ 
in the entire sample, $\epsilon_\cat$ is the 
fraction of signal events tagged in category $\cat$, and
$N^{\cont, h}_\cat$ is the number of continuum background 
events with bachelor track of type~$h$ that are tagged in 
category~$\cat$. The total likelihood ${\cal L}$ is the product 
of likelihoods for each tagging category.

The probability density functions (PDFs)   
${\cal P}_{\cat}^{\rho h}$,  ${\cal P}_{\cat}^{\cont, h}$ 
and the likelihood  terms ${\cal L}^{\B,h}_{j,\cat}$ are the
product of the PDFs of five discriminating variables.
The signal PDF is thus given by
${\cal P}_\cat^{\rho h} = 
	{\cal P}^{\rho h}(\mes)\cdot 
	{\cal P}^{\rho h}(\de) \cdot 
	{\cal P}^{\rho h}(\NN) \cdot 
	{\cal P}^{\rho h} (\theta_{C})
	\cdot {\cal P}_\cat^{\rho h}(\deltat)$, 
where 
${\cal P}_\cat^{\rho h}(\deltat)$ contains the measured physics 
quantities defined in Eq.~(\ref{eq:thTime}), diluted by the effects of 
mistagging and the~$\deltat$ resolution. The PDF of the continuum 
contribution with  bachelor track $h$ is denoted 
${\cal P}_{\cat}^{\cont, h}$. 
The likelihood term ${\cal L}^{\B,h}_{j,\cat}$ corresponds to 
the \B-background contribution $j$ of the $N_B$ 
categories. 

The signal PDFs are decomposed into three parts 
with distinct distributions: 
signal events that are correctly reconstructed, misreconstructed signal 
events with right-sign $\rho$ charge, and misreconstructed signal events 
with wrong-sign $\rho$ charge. Their individual fractions are estimated
by MC simulation. 
The $\mes$, $\de$, and \NN\  output PDFs for signal
and $B$ background are taken from the simulation except for the
means of the signal Gaussian PDFs for $\mes$ and $\de$,  which are free 
to vary in the fit.
The continuum PDFs are described by six free parameters.
The $\theta_C$ PDF is modeled as in Ref.~\cite{bib:BabarSin2alpha}.  
The $\deltat$-resolution function for signal and $B$-background 
events is a sum of three Gaussian distributions, 
with parameters determined 
from a fit to fully reconstructed $\Bz$ decays~\cite{bib:BabarS2b}.
The continuum $\deltat$ distribution is parameterized as the sum of 
three Gaussian distributions with common mean, two relative fractions,
and three distinct widths that scale the $\dt$ event-by-event error, 
yielding six free parameters.
For continuum, two charge asymmetries and the ten parameters 
$N^{\cont, h}_\cat$ 
are free. A total of 34 parameters, including signal yields and the
parameters from Eq.~(\ref{eq:thTime}), are varied in the fit.

\begin{table}[t]
\caption{Summary of the systematic uncertainties.}
{\small
\begin{center}
\setlength{\tabcolsep}{0.165pc}
\begin{tabular}{lcccccccc}
\hline
\hline
 & $N^{\rho K}$ & $N^{\rho\pi}$
 & $\AcprhoK$ & $\Acprhopi$ & $C_{\rho\pi}$  & $\dC_{\rho\pi}$ 
& $S_{\rho\pi}$ & $\dS_{\rho\pi}$  \\
\rs{Error source} & \mc{2}{c}{(events)}
 &  \mc{6}{c}{(in units of $10^{-2}$)} \\
\hline
$\deltamd$ and $\tau$ & 0.1 & 0.1  &
0.0 & 0.0 & 0.4 & 0.4 & 0.2 & 0.1  \\

$\dt$ PDF & 1.2  & 1.9  &
 0.4 & 0.2 & 1.4 & 0.8 & 1.5 & 1.2   \\

Signal model & 4.0 & 13.1  &
 1.2 & 0.8 & 0.7 & 0.8 & 1.4 & 1.0   \\

Particle ID &  0.6 &  0.7 &
0.5 & 0.2 & 0.1 & 0.1 & 0.1 & 0.1  \\

Fit procedure & 8.0  & 15.7 &
 0.4 & 0.2 & 0.4 & 0.4 & 0.4 & 0.3  \\

DCS decays & 0.0  & 0.3 &
0.0  & 0.1 & 2.2 & 2.2 & 0.8 & 0.7  \\

$B$ background & 16.0  & 14.2  &
7.9 & 2.8 & 3.0 & 3.5 & 2.1 & 1.8  \\
\hline
Total  & 18.4  &  25.0 &
8.0 & 2.9 & 4.1 & 4.3 & 3.1  & 2.5  \\
\hline
\hline
\end{tabular}
\end{center}
}
\label{tab:sys_table}
\end{table}
The contributions
to the systematic error on the signal 
parameters are summarized in Table~\ref{tab:sys_table}. 
The uncertainties associated with \dm\ and $\tau$ 
are estimated by varying these parameters within
the uncertainty on the world average~\cite{PDG2002}.
The uncertainties due to the signal model 
are obtained from a control sample of fully reconstructed 
$B^{0} \rightarrow D^{-} \rho^{+}$ 
decays. 
We perform fits on large MC samples with the measured proportions
of $\rhopi/\rhok$ signal, and continuum and $B$ backgrounds.
Biases observed in these tests are due to imperfections in the 
PDF model; \eg, unaccounted correlations 
between the discriminating variables of the signal and $B$-background 
PDFs. The biases are added in quadrature and 
assigned as a systematic uncertainty of the fit procedure.
The systematic errors  due to interference between the 
doubly-Cabibbo-suppressed 
(DCS) $\bar b \to \bar u c \bar d$ amplitude with the Cabibbo-favored 
$b \to c \bar u d$ amplitude for tag-side $B$ decays
have been estimated from simulation 
by varying freely all relevant strong phases~\cite{bib:DCSD2003}.

The main source of systematic uncertainty is
the $B$-background model. 
The expected event yields from the background modes 
are varied according to the uncertainties in the measured or estimated 
branching fractions.
Systematic errors 
due to possible nonresonant $B^0\to\pi^+ \pi^-\pi^0$ decays
are derived from experimental limits~\cite{bib:BaBarRhopi}.
Repeating the fit without using the 
$\rho$-candidate mass and helicity angle gives results that 
are compatible with those reported here.
Since $B$-background modes 
may exhibit 
\CP violation, 
the corresponding parameters are varied within
their physical ranges.

\begin{figure}[th]
  \centerline{ 	\epsfysize4.2cm\epsffile{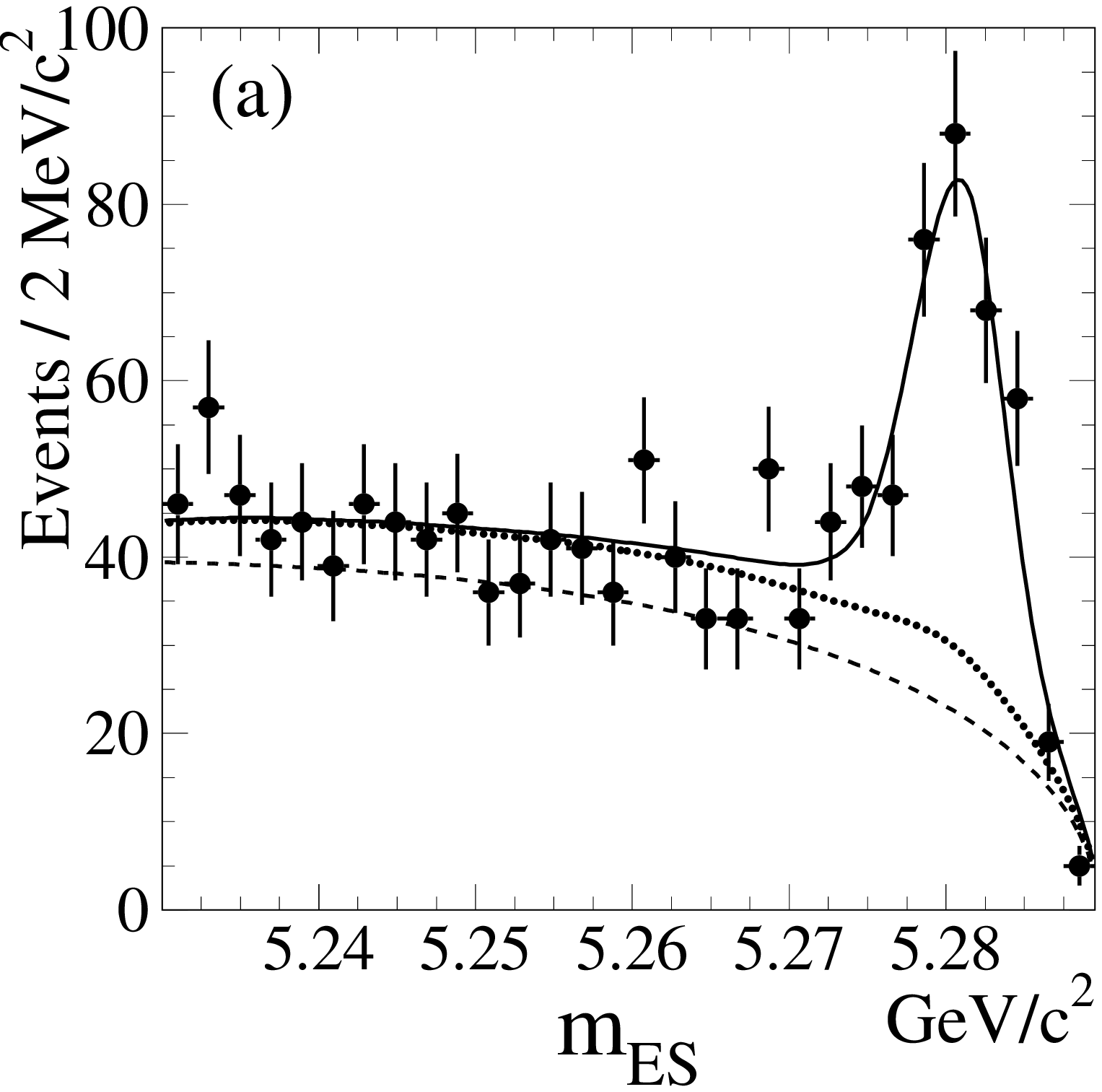}
		\epsfysize4.2cm\epsffile{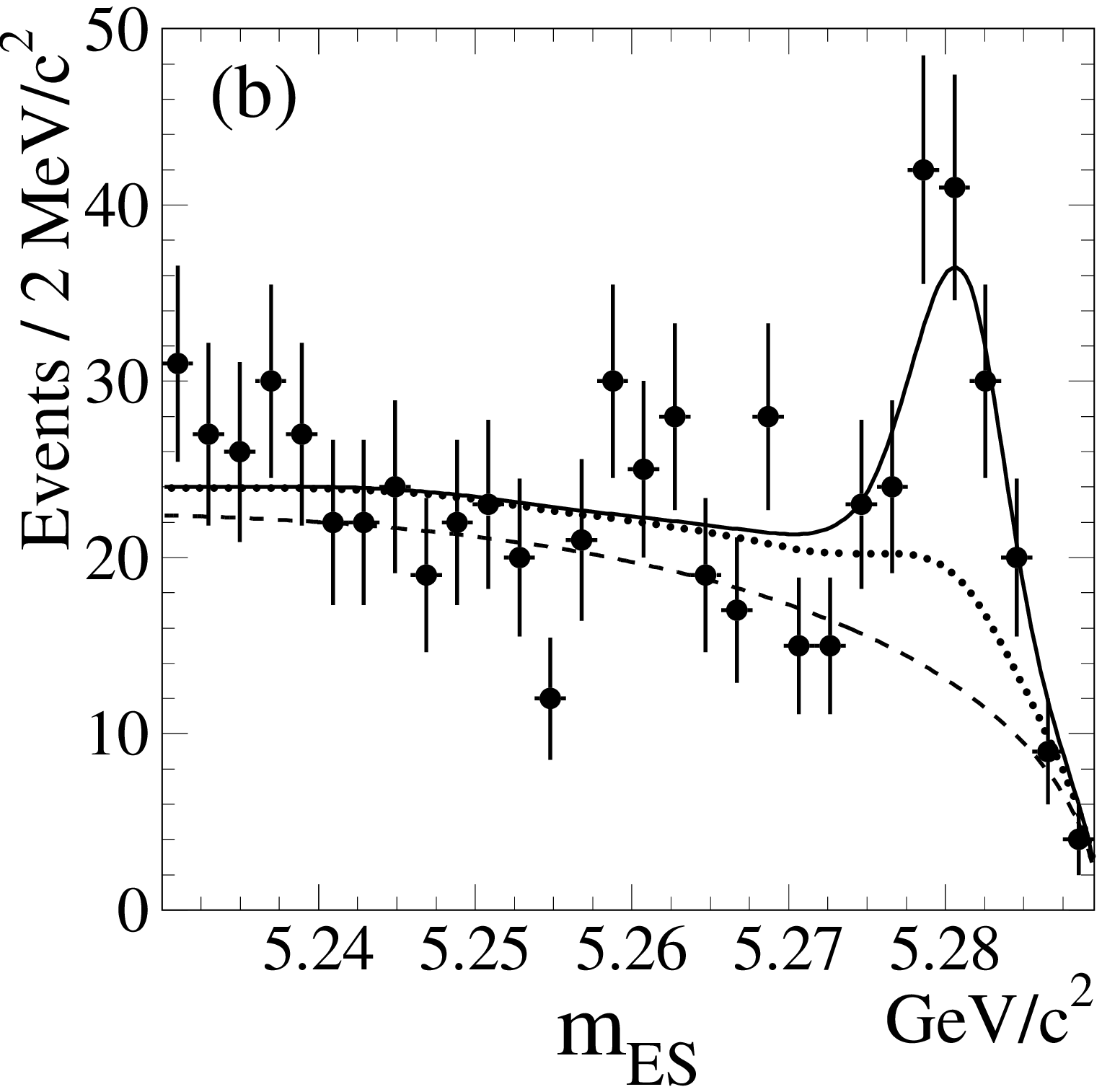}}
  \centerline{ 	\epsfysize4.2cm\epsffile{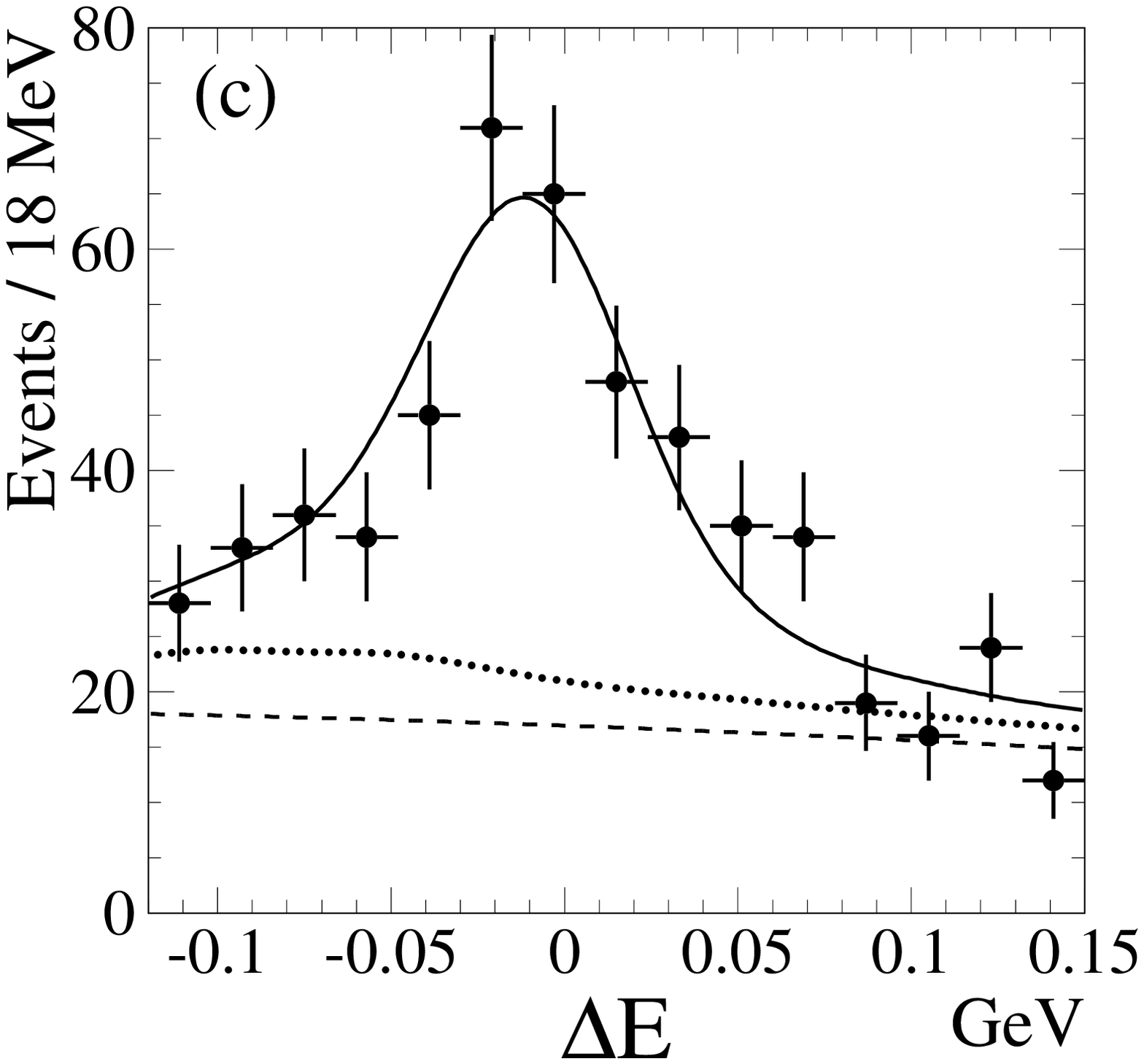}
                \epsfysize4.2cm\epsffile{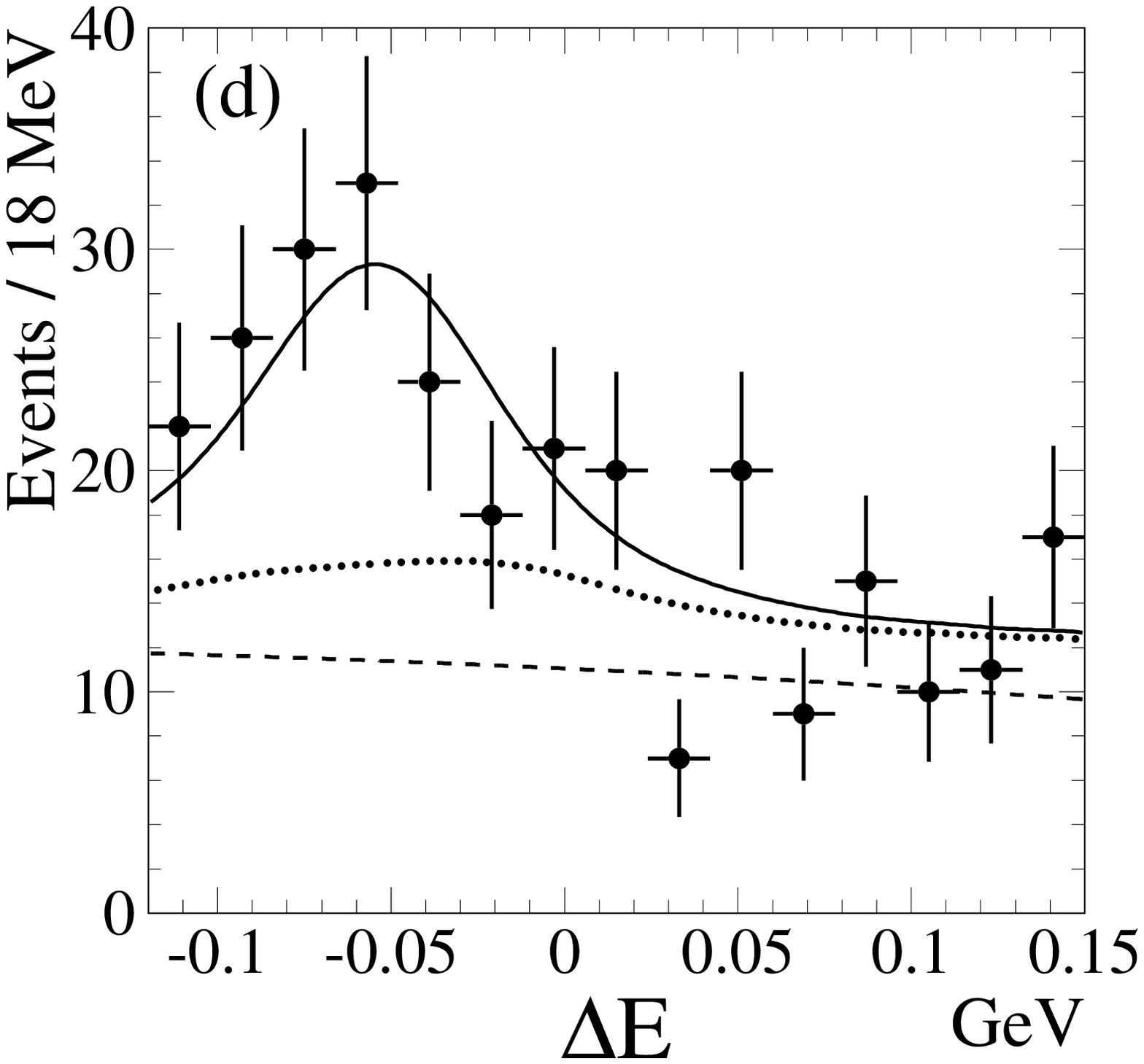}} 
\vspace{-0.3cm}
\caption{Distributions of $\mes$ and $\de$ for samples enhanced 
	in $\rho\pi$ signal (a,c) and $\rho K$ signal (b,d).
	The solid curve represents a projection of the maximum 
	likelihood fit result. The dashed curve represents the contribution 
	from continuum events, and the dotted line indicates the combined 
	contributions from continuum events and $B$-related backgrounds.}
\label{fig:ProjMesDE}
\end{figure}
\begin{figure}[th]
  \begin{center}
    \includegraphics[width=0.43\textwidth,angle=-90]{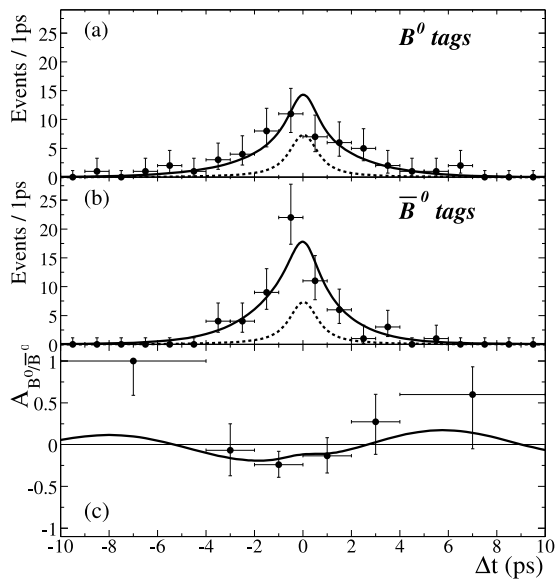}
  \end{center}
\vspace{-0.3cm}
\caption{Time distributions for events
	selected to enhance the $\rho\pi$ signal tagged as
	(a) $\Bz_{\rm tag}$ and (b) $\Bzb_{\rm tag}$, and (c) 
	time-dependent asymmetry
	between $\Bz_{\rm tag}$ and $\Bzb_{\rm tag}$. The solid curve
	is a likelihood projection of the fit result.
	The dashed line is the total \B- and continuum-background 
	contribution. }
	\label{fig:asymCS} 
\end{figure}

The maximum likelihood fit results in the event yields
$N^{\rho\pi} = 428^{\,+34}_{\,-33}$ and $N^{\rho K} = 120^{\,+21}_{\,-20}$,
where the errors are statistical. Correcting the yields  by a small fit 
bias determined using the MC simulation ($3\%$ for $\rho\pi$ and
$0\%$ for $\rho K$), we find for the branching fractions
\begin{eqnarray*}
  {\cal B}(\Bz\to\rho^{\pm}\pi^{\mp}) &=& 
	(22.6 \pm 1.8 
	\pm2.2)\times 10^{-6}\,, \\
  {\cal B}(\Bz\to\rho^{-}K^{+}) &=& 
	(7.3^{\,+1.3}_{\,-1.2}
	\pm1.3)\times 10^{-6}\,,
\end{eqnarray*}
where the first errors are statistical and the second systematic.
The systematic errors include an uncertainty of $7.7\%$ for efficiency 
corrections, dominated by the uncertainty in the $\pi^0$ reconstruction 
efficiency.
Figure~\ref{fig:ProjMesDE} shows distributions of $\mes$ and $\de$, 
enhanced in signal content by cuts on the signal-to-continuum 
likelihood ratios of the other discriminating variables. 
For the \CP-violating parameters, we obtain
\begin{eqnarray*}
	\Acprhopi     =  - 0.18\pm 0.08 \pm{0.03}\,,&   
	\AcprhoK      = 0.28 \pm 0.17 \pm 0.08\,,\\
	C_{\rho\pi}   = \phantom{-}0.36 \pm 0.18 \pm 0.04\,,& 
	\,S_{\rho\pi} =  0.19\pm 0.24\pm 0.03\,.
\end{eqnarray*}
For the other parameters in the description of the $\Bz(\Bzb) \to \rhopi$
decay-time dependence, we find
\begin{eqnarray*}
	\dC_{\rho\pi} = 0.28^{+0.18}_{-0.19}\pm 0.04\,, & 
	\dS_{\rho\pi} = 0.15\pm 0.25\pm 0.03~.
\end{eqnarray*}
We find the linear correlation coefficients 
$\corr_{C,\dC}=0.18$
and
$\corr_{S,\dS}=0.23$, while all 
other correlations are smaller.
As a validation of our treatment of the time dependence 
we allow
$\tau$ and $\deltamd$ to vary in the fit. We find $\tau = (1.64\pm 0.13)\ps$ 
and $\deltamd = (0.52\pm 0.12)\ps^{-1}$; the remaining free parameters 
are consistent with the nominal fit.
The raw time-dependent 
asymmetry $A_{\Bz/\Bzb}=  (N_{\Bz} - N_{\Bzb})/(N_{\Bz} + N_{\Bzb})$ 
in the tagging  categories  dominated by kaons and leptons is 
represented in Fig.~\ref{fig:asymCS}.

In summary, we have presented measurements of branching fractions and 
\CP-violating asymmetries in $\Bz\to\rho^\pm\pi^\mp$ and $\rho^-K^+$ 
decays. We do not observe direct or mixing-induced \CP  violation 
in the time-dependent asymmetry of $\Bz\to\rho^\pm\pi^\mp$ decays and
there is no evidence for direct \CP  violation in $\Bz\to\rho^-K^+$.

\par

We are grateful for the excellent luminosity and machine conditions
provided by our \pep2\ colleagues, 
and for the substantial dedicated effort from
the computing organizations that support \babar.
The collaborating institutions wish to thank 
SLAC for its support and kind hospitality. 
This work is supported by
DOE
and NSF (USA),
NSERC (Canada),
IHEP (China),
CEA and
CNRS-IN2P3
(France),
BMBF and DFG
(Germany),
INFN (Italy),
FOM (The Netherlands),
NFR (Norway),
MIST (Russia), and
PPARC (United Kingdom). 
Individuals have received support from the 
A.~P.~Sloan Foundation, 
Research Corporation,
and Alexander von Humboldt Foundation.

\end{document}